# Work from Home and Job Satisfaction:
# Differences by Disability Status among Healthcare Workers

*Yana van der Meulen Rodgers, PhD*
*Rutgers University, 94 Rockafeller Road, Piscataway, NJ  08854 USA*

*Lisa Schur, PhD*
*Rutgers University, 94 Rockafeller Road, Piscataway, NJ  08854 USA*

*Flora M. Hammond, MD*
*Indiana University School of Medicine, 340 West 10th Street, Indianapolis, IN 46202 USA*

*Renee Edwards, PhD*
*Rutgers University, 30 Livingston Avenue, New Brunswick, NJ 08901, USA*

*Jennifer Cohen, PhD*
*Rutgers University, 94 Rockafeller Road, Piscataway, NJ  08854 USA*

*Douglas Kruse, PhD*
*Rutgers University, 94 Rockafeller Road, New Brunswick, NJ  08854 USA*

**Abstract**
**Background:** Many workers with disabilities face negative stereotypical attitudes, pay gaps, and a lack of respect in the workplace, contributing to substantially lower job satisfaction compared to people without disabilities. Work from home may help to increase job satisfaction for people with disabilities.
**Objective:** This study analyzes how different measures of job satisfaction vary between people with and without disabilities, and the extent to which working from home moderates the relationship between disability and job satisfaction.
**Methods:** We use multivariable regression analysis to examine if the ability to work from home moderates the relationship between disability and indicators related to job satisfaction. The dataset draws on a novel survey of healthcare professionals.
**Results:** Results show that people with disabilities have relatively greater turnover intentions, lower sense of organizational commitment and support, weaker perceptions of openness and inclusion in the workplace, and worse relations with management and coworkers. Regressions indicate that working from home helps to improve most perceptions of work experiences but does so more for people without disabilities than for people with disabilities.
**Conclusions:** The findings suggest that (a) some accommodations typically viewed as exceptions to meet the needs of people with disabilities have even greater benefits for the workforce at large and (b) because workers with and without disabilities benefit from remote work, we cannot expect those accommodations to close the gaps caused by inequities.

**Corresponding Author:** Yana van der Meulen Rodgers, 94 Rockafeller Road, Piscataway, NJ, 08854, USA. Tel 848-932-4614. Email yana.rodgers@rutgers.edu

**Acknowledgments:** The authors acknowledge So Ri Park for her valuable research assistance. We also thank participants at the West Virginia University Center for Excellence in Disabilities seminar, the National Association for Forensic Economics session at the 2025 ASSA conference, and the Doctorado de Ciencias Sociales seminar at Universidad de Buenos Aires.

**Funding Statement:** This research received funding support from the National Institute on Disability, Independent Living, and Rehabilitation Research (NIDILRR) for the Rehabilitation Research & Training Center (RRTC) on Employment Policy: Center for Disability-Inclusive Employment Policy Research Grant [grant number #90RTEM0006-01–00] and the NIDILRR RRTC on Employer Practices Leading to Successful Employment Outcomes Among People with Disabilities Research Grant [grant number #90RTEM0008-01-00].

**Data and Programs:** For information regarding the data and/or computer programs used for this study, please contact the first author at yana.rodgers@rutgers.edu.

# 1. Introduction

Many workers with disabilities face negative stereotypical attitudes and expectations from managers and coworkers that limit their career growth and the quality of their work-life balance.[1,2] They also face a gap in pay after accounting for productive characteristics such as education and job experience.[2-4] These disparities are linked to lower job satisfaction among workers with disabilities.[2] This job satisfaction gap between workers with and without disabilities is well established in the literature and has also been linked to a lack of respect for people with disabilities: U.S. workers with disabilities have half the odds of reporting high job satisfaction compared to workers without disabilities, with workplace respect explaining 38% of the gap.[5]

Workplace accommodations can help to improve job performance and satisfaction,[6,7] and work from home can be a reasonable accommodation. However, employers were generally resistant to work from home before the COVID-19 pandemic, and courts did not consider it a reasonable accommodation in most cases under the Americans with Disabilities Act.[8,9] Such resistance declined when the pandemic forced employers to reconsider the best ways to accomplish job tasks during the pandemic, and the opportunity to work from home was extended to many people with and without disabilities. Notably, while workers with disabilities were more likely than those without disabilities to work from home before the pandemic, they were less likely to work from home during the pandemic, largely due to differences in the occupational distribution. [10,11]

Given the increased acceptance of work from home among employers, at least during the pandemic, we consider the extent to which work from home can narrow the job satisfaction gap between people with and without disabilities. We address this question using a dataset drawn from a novel survey of healthcare workers from a large employer in the healthcare sector. Our analysis tests three key hypotheses: (1) individuals with disabilities report lower levels of job satisfaction and have less favorable work experiences compared to their counterparts without disabilities; (2) work from home improves job satisfaction and work experiences for all workers; and (3) work from home helps narrow the job satisfaction gap between people with and without disabilities. We consider the impact of work from home on a direct measure of job satisfaction and ten different indicators of perceptions of work experiences that contribute to job satisfaction, including job autonomy, the employer's treatment of people with disabilities, and engagement with coworkers and management.

The healthcare sector is the focus of this analysis for three reasons. First, the industry employs a relatively high proportion of individuals with disabilities, particularly in entry-level roles. While 4.9% of the total labor force comprises people with disabilities, they make up 5.5% of healthcare support staff.[12] Second, the healthcare sector provides an opportunity to study a diverse workforce, as it disproportionately employs underrepresented men and women of various races, ages, and educational backgrounds across various roles. This diversity may extend the applicability of our findings to other sectors of the economy. Finally, the overall health of the population is closely tied to the resources available to and the capacity of the healthcare system, rendering it critically important to social wellbeing.

## 2. Methods

We collaborated with a large state-wide university-affiliated health system to design and administer an original survey. The survey targeted healthcare employees with and without disabilities, concentrating on their experiences with employer policies regarding remote work. Between May 26 and July 31, 2023, 1,405 employees participated in the survey, which we conducted using the Qualtrics platform. The survey instrument was motivated by Schur et al. (2014; 2020) and incorporated scales frequently used in organizational behavior studies.[13,14] The online Appendix describes these scales. After excluding responses with missing information, the final dataset comprised 993 participants. Sample means for all variables are reported in Appendix Tables 1 and 2. This project was approved by the XXX Institutional Review Board (ID Pro2021002068), and all respondents provided their informed consent.

We used the data to calculate summary statistics on disability status, work from home, and perceptions of workplace experiences related to job satisfaction, such as inclusiveness, treatment of people with disabilities, and relationships with managers and coworkers. We then estimated multiple regressions to examine if the ability to work from home moderates the relationship between disability and job satisfaction, controlling for other characteristics. The complete model specification is:

$$Outcome_i = b_0 + b_1 Disab_i + b_2 WFH_i + b_3 Disab_i * WFH_i + b_4 X_i + e_i \ .$$

The notation $Outcome_i$ is job satisfaction and a set of ten indicators of workplace experiences for person i, $Disab_i$ denotes a dummy variable for disability status, and $WFH_i$ measures the extent to which someone could work from home. We model $WFH_i$ as a set of dummy variables based on the frequency of working from home: not at all, less than three days per week, and three or more days per week. In a secondary set of results found in the online Appendix, we also model $WFH_i$ based on the reason for working from home: not at all, for the pandemic or the employer's benefit, and for the employee's benefit. The notation $X_i$ is a set of control variables including age, gender, race/ethnicity, marital status, education, income above $75,000, number of children at home, managerial role, full-time worker, and tenure at the employer.

Job satisfaction is measured directly by a survey question, and workplace experiences are measured by 10 indices constructed from survey questions about job autonomy,[15-17] turnover intentions,[18] organizational commitment,[19] organizational citizenship behaviors,[20] perceived organizational support,[21-22] employer openness to differences, the climate for inclusion,[23] treatment of people with disabilities, relationship with one's manager (also known as leader-member exchange),[24] and relationships with one's coworkers (also known as coworker exchange).[25]

Disability is assessed using the six standard questions from the Census, supplemented by two additional questions regarding challenges in social interactions and long-term activity limitations. The eight questions are: (1) "Are you deaf or do you have serious difficulty hearing?"; (2) "Are you blind or do you have serious difficulty seeing even when wearing glasses?"; (3) "Do you have serious difficulty concentrating, remembering, or making decisions?"; (4) "Do you have serious difficulty walking or climbing stairs?"; (5) "Do you have difficulty dressing or bathing?"; (6) "Do you have difficulty doing errands alone such as visiting a doctor's office or shopping?"; (7) "Do you have difficulty interacting and/or communicating with others?"; and (8) "Do you have a long-term health problem or impairment that limits the kind or amount of work, housework, school, parenting, recreation, or other activities you can do?" These questions were not mutually exclusive, and respondents could choose more than one. An individual is classified as having a disability if they responded affirmatively to any of these questions.

We may underestimate the actual number of individuals with disabilities. The Census questions are designed to capture functional disabilities with more serious levels of limitations.[26] The questions may not capture mental health, chronic pain, or people who are neurodivergent and have challenges related to what have historically been called learning disabilities like ADHD, autism, and a range of other atypicalities. Therefore, our survey includes two additional questions on long-term impairments and challenges in social interactions in an effort to capture some of these people. Still, some individuals may hesitate to indicate in a survey that they have difficulty with activities or social interactions due to the persistent stigma surrounding disability.

Further, estimates of accommodation requests for a disability are inherently restricted to those who are aware of or have been diagnosed with a disability, potentially excluding those who remain undiagnosed or unaware. Even individuals with a diagnosed disability who wish to request remote work accommodations may be discouraged by the requirement to provide medical documentation, especially if they are not currently undergoing treatment for their condition.

Because disclosure allows employers to accommodate employee needs and help create an equitable opportunity to succeed, workers who disclose may have higher job satisfaction. Hence, our analyses use two alternative constructions of the disability sample: (1) those who reported one or more of the disability conditions in the survey (n=228), and (2) those who have disclosed a disability to their employers (n=114). In both cases, the total analytic sample is 993 individuals. Results for people who reported a disability in the survey are reported in this paper, and results for people who disclosed a disability to their employer are reported in the online Appendix and are not reported here due to space limitations.

## 3. Results

*3.1 Descriptive statistics for disability or health impairment*

Table 1 shows that of the 993 respondents in our sample, 23.0% report functional limitations, challenges in social interactions, and/or a long-term impairment. We refer to this response as having a disability or health impairment. This percentage drops to 18.4% if we restrict our measure to the six kinds of disability reported in Census data, almost double the 9.4% estimate based on data from the 2024 American Community Survey (ACS) for the percentage of healthcare workers with a disability in the organization's state. One explanation is that we may have over-sampled people willing to report a disability or health impairment. Timing may also have played a role, with our 2023 survey picking up more respondents with functional disabilities related to COVID-19 compared to the 2024 ACS. Accordingly, the high rate of disability yields sufficient sample sizes of people living with at least one disability and people without a disability to conduct comparative analyses. Among people with disabilities, close to half (46.5%) reported a long-term impairment. The most common type of functional disability was difficulty concentrating and making decisions, followed by difficulty walking and climbing stairs. Of the group with a disability, about one-half (51.8%) disclosed their disability to their employer, another quarter (25.2%) did not disclose, and the remainder said that it is "complicated" (23.0%) or did not respond (0.9%). Reluctance to disclose is usually widespread due to the stigma associated with having a disability.[27-30]

*Table 1 Here*

We next establish a gap in job satisfaction and perceptions of job experiences by disability status, which paves the way to examining whether working from home equalizes those perceptions.

*3.2 Descriptive statistics for job satisfaction and work experiences*

As shown in Table 2, there is no statistically significant difference in job satisfaction between people with and without disabilities as measured by the single survey question on job satisfaction. However, Table 2 provides compelling evidence of relatively more negative perceptions of workplace experiences that contribute to job satisfaction among people with disabilities. The table reports sample means for this study's 10 indices of perceptions of work experiences, scaled between zero and one and constructed from different sets of questions in the survey (see Appendix Table 3).

*Table 2 Here*

*Turnover intentions*

People with disabilities had a higher score for the turnover intention index than those without disabilities (0.420 versus 0.323, p<0.01). Examining the specific items in the turnover intentions measure, employees with disabilities have a relatively higher likelihood of planning to look for a job outside of the organization, thinking often of quitting their job, and desiring a new job. These differences, though, were not as large for people who had disclosed their disabilities to their employer (0.407 versus 0.337, p<0.10).

*Organizational commitment, support, and citizenship behaviors*

Closely related to turnover intentions, people with disabilities had a lower index of organizational commitment compared to people without disabilities (0.459 versus 0.530, p<0.05). Driving this result was a lower likelihood of people with disabilities saying that they feel a strong sense of belonging at the employer and that they feel like they are a "part of the family." While people with and without disabilities have similar responses to questions about organizational citizenship behaviors, people with disabilities believe less strongly that they are supported by their organizations (0.309 versus 0.403, p<0.01). Comprising this index of perceived organizational support are perceptions that the employer cares about their wellbeing and opinions, and that the employer takes pride in their accomplishments at work.

*Workplace inclusiveness and openness to difference*

Perceptions of inclusiveness in the workplace mirror the negative relationship between disability and perceived organizational support. Table 2 shows that people with disabilities are substantially less likely than people without disabilities to believe that their employer is open to differences (0.453 versus 0.561, $p<0.01$). Underlying this index are questions about whether people can reveal their true selves at work, whether employees are valued as people rather than merely for their jobs, and whether the work culture appreciates the differences that people bring to the workplace. People with disabilities are also relatively less likely to believe that the employer has an inclusive workplace climate (0.348 versus 0.416, $p<0.05$), with more skepticism that the employer actively seeks employee input, uses employee insights to redefine work practices, and considers input from people in different roles and functions when problem-solving. For the perceptions of openness to differences and of inclusiveness, the disability gaps become substantially smaller and lose their statistical significance when we consider disclosed disabilities.

*Treatment of people with disabilities*

Table 2 further shows that there is no substantial difference between people with and without disabilities in the index of perceptions on how people with disabilities are treated. However, this aggregate index masks some differences among the more detailed questions that are reported in Appendix Table 3. People with disabilities are more likely to state that their workplace has a bias against people with disabilities, and that employees without disabilities are treated better than employees with disabilities. People without disabilities tend to have a more favorable view of the culture around disability at their workplace, being more likely to agree that employees treat people with disabilities with respect and that their manager is responsive to the needs of people with disabilities.

*Workplace relationships*

Some of the biggest and most robust differences between people with and without disabilities appear in Table 2's results for relationships with supervisors and coworkers. People with disabilities are uniformly less likely than people without disabilities to agree to various descriptors of a positive relationship with one's manager, including knowing how satisfied the manager is with one's performance, feeling that the manager is understanding, feeling that the manager recognizes one's potential, being able to count on the manager for support during a tough situation, having an effective working relationship with one's manager, and believing that the manager would use their power and influence to help the employee. These disability differences are mirrored in the responses for the relationship with one's coworkers.

In sum, the results in Table 2 show that people with disabilities have relatively greater turnover intentions, a weaker commitment to their employer, a lower sense of organizational support, weaker perceptions of an open and inclusive workplace, and worse relations with management and coworkers. Each of these measures of workplace experiences can contribute to lower overall job satisfaction for people with disabilities. As shown in Table 2's final column, which reports regression-adjusted disability gaps, these statistically significant disparities between people with and without disabilities still hold even after controlling for differences in education and other observed characteristics. Notably, people with disabilities exhibit the same level of citizenship behaviors as those without disabilities despite lower organizational support, weaker perceptions of an open and inclusive workplace, and worse relations with management and coworkers.

*3.3 Descriptive statistics for work from home*

The results on work from home in Table 3 are mostly similar between people with and without disabilities. Before the pandemic, individuals with and without disabilities had similar likelihoods of working from home (0.189 versus 0.160; p>0.10), and at the time of the survey, the likelihood for both groups of working from home was the same (0.529). This result differs from earlier research indicating that people with disabilities were more likely to work from home before the pandemic and less likely to do so during the pandemic.[11, 31] A possible explanation for the difference is that in our survey, one of the more common occupations among respondents both with and without disabilities is administrative support staff, while the earlier research used a nationally representative sample of workers.[11, 31]

*Table 3 Here*

We aggregated several survey questions about work from home to create an index of agreement that working from home before the pandemic had positive effects, and an index of agreement that currently working from home has positive effects. As shown in the last two rows of Table 3, the mean values of these indices are similar between people with and without disabilities. That said, among people who responded that they currently work from home, people with disabilities are more likely than people without disabilities to work at least three days per week at home (0.825 versus 0.716, p<0.05). Importantly, people with disabilities are more likely to desire more work from home (0.462 versus 0.333, p<0.05), indicating that there is unmet demand for remote work from people with disabilities.

People who have disclosed their disabilities to the employer have stronger results for the positive assessment of work from home. As shown in the online Appendix, the index of agreement on work from home having positive effects is substantially higher for people with disclosed disabilities compared to people with no or undisclosed disabilities (0.720 versus 0.623,

p<0.05). Among the detailed indicators used to construct the index of agreement on work from home having positive effects, the positive assessment of work from home among people with disclosed disabilities is especially true for morale, staying at the employer, relations with coworkers, and productivity.

Given that our results show more negative perceptions of work experiences among people with disabilities compared to people without disabilities, we now consider whether working from home helps to narrow these gaps.

*3.4 Regression analysis*

Regression results for the association between job satisfaction, disability status, and work from home are found in Table 4. The table reports results for the single job satisfaction question and then for each of the 10 indices of perceptions of work experiences. Recall that all regressions include a set of dummy variables for the frequency of working from home (not at all, less than three days per week, and three or more days per week) and their interaction with disability status. The regressions do not include a disability main effect, so the coefficients for the interactions between disability and work from home represent the full disability effect within each frequency category.

*Table 4 Here*

Results for the baseline effects of work from home frequency (regardless of disability status) are positive and statistically significant for one or both of the work from home variables in the case of job satisfaction and in eight of the ten measures of work experiences. For example, job satisfaction is significantly higher among those working at home up to two days a week (0.124) or three or more days a week (0.195), both significant at $p<0.01$, relative to those doing no work from home. Across regressions, the coefficient magnitude on work from home is consistently larger for three or more days per week compared to under three days. Working from home is also favorable for reducing turnover intentions: turnover intentions are significantly lower among those working at home three or more days a week (-0.162, p<0.01) compared to those doing no work from home. Work from home has no statistically significant association with perceptions of relationships with coworkers.

While the effects of work from home are positive for the full sample, the effects are often less positive for employees with disabilities. These interaction term coefficients must be interpreted in the context of the positive baseline effect. Most of the disability interactions are negative, and many are statistically significant, indicating less favorable effects of working from home relative to people without disabilities. Note that this does not mean working from home negatively affects perceptions of job experiences for employees with disabilities—only that the effects are not as positive. For example, the organizational commitment regression shows a positive base effect of 0.176 for working from home three or more days a week, and the disability interaction shows a coefficient of -0.117, indicating that the full effect of working from home for employees with disabilities is 0.176−0.117 = 0.059. Hence the organizational commitment of employees with disabilities who work three or more days a week is 0.059 higher than that of on-site employees with disabilities. This net positive effect for people with disabilities is smaller than the 0.176 difference among employees without disabilities.

The finding that work from home (for one or both frequencies) has a less positive effect for employees with disabilities than for those without disabilities holds for job autonomy, organizational commitment, perceived organizational support, employer openness to difference, and climate for inclusion. This pattern also holds for turnover intentions, where working from home three or more days a week is linked to lower turnover intentions for employees without disabilities (base effect = -0.162, p<0.01), while the disability interaction indicates a weaker relationship for people with disabilities; work from home may, however, be especially important for employees with disabilities given that not working from home is strongly linked to greater turnover intentions among employees with disabilities (0.159, p<0.01). The only measures where working from home has a net negative effect on employees with disabilities are supervisor and coworker relations, where the negative disability interactions exceed the positive base effects. For example, the managerial relations regression shows a positive base effect of 0.124 for working from home three or more days a week, and the disability interaction shows a coefficient of -0.171, indicating that the full effect of working from home for employees with disabilities is 0.124−0.171 = -0.47.

Overall, we find that although individuals with disabilities do not report significantly lower levels of job satisfaction as measured by a single question, those with disabilities do have less favorable work experiences compared to their counterparts without disabilities (H1). Work from home improves job satisfaction and work experiences for all workers (H2), but fails to narrow the job satisfaction gap as measured by work experiences between people with and without disabilities (H3). In most cases, working from home has positive effects for people with disabilities, especially for those who can work three or more days from home. Still, it does not narrow the disability gaps in work experiences because the gains of working from home are even more substantial for people *without* disabilities.

## 4. Discussion

This study has found a substantial disability gap in perceptions of work experiences, where people with disabilities report greater turnover intentions, lower sense of belonging/inclusion, worse relations with management and coworkers, and weaker perceptions of openness and inclusion. These findings are consistent with an earlier study using national data from the General Social Survey.[2] Our regression results indicate that working from home has a positive effect on many perceptions of work experiences for employees with disabilities but does not close disability gaps in those perceptions. The reason is that working from home has an even stronger positive effect for employees without disabilities. The weaker positive effects among people with disabilities appear to be tied to social isolation, as employees with disabilities who work from home report especially poor supervisor and coworker relations. The finding that supervisor and coworker relations are especially poor among employees with disabilities who work from home suggests that such work can reinforce the social isolation of employees with disabilities since it may involve fewer social interactions, less visibility ("out of sight, out of mind"), fewer opportunities for networking, communication problems, inadequate access to resources, and lack of trust. Working from home may also aggravate problems with work boundaries, work-life balance, and distractions at home for employees with disabilities.

Work from home is generally associated with more positive job engagement and better job experiences among employees with and without disabilities. Critically, then, our finding of smaller positive work from home effects among employees with disabilities means that to narrow disability gaps in job satisfaction, employers must explore policies and programs that disproportionately benefit those with disabilities. Further, employers should reduce the social isolation of employees with disabilities working at home. Such a strategy could include more consistently scheduled online supervisor and coworker interaction, virtual social events, and occasional accessible in-person events. The predictability of these interactions is important - both their timing and content - so that people can prepare themselves. Such efforts must reflect an awareness that employees have different kinds of disabilities, which would affect the feasibility, cost, and comfort level of increased social interaction in different ways for employees with disabilities. Improving the work experiences of people with disabilities ought to be a regular business practice, not an exception for individuals who become characterized as deviations from the norm.

## 5. Conclusion

Future research could provide deeper insights by focusing on ways to increase access to meaningful employment for individuals with disabilities and improve their workplace experiences after securing a job. Another fruitful area of research revolves around the accessibility and design of remote work tools. One could evaluate the accessibility of popular remote work platforms and tools for people with different types of disabilities, or investigate how emerging technologies like artificial intelligence, virtual reality, and augmented reality can enhance accessibility for remote workers with disabilities. Another area of future research stemming from our findings is the mental health benefits of remote work for people with disabilities, including reduced stress from commuting and increased comfort in home environments. At the same time, such research must assess the potential challenges, such as isolation or lack of social support, and propose interventions to mitigate them.

The finding that work from home strongly benefits people without disabilities is consistent with studies showing that policies that benefit marginalized groups often benefit dominant groups more.[32-34] This does not mean those policies should be avoided; they should be part of universal design applied to the workplace. However, we should not expect those policies to eliminate inequities. In the case of disability, one of the reasons the dominant group benefits is that the line between groups with and without disabilities is not sharply defined. Many people do not know they have disabilities. Those who do know may not think they require accommodations, and requesting accommodations remains stigmatized for those who could benefit from them. Closing gaps will require interventions that disproportionately benefit people seeking accommodation for disabilities – interventions beyond those that improve working conditions more broadly.

Table 1. Descriptive Statistics for Disability Status

|  | Number | Percent |
|---|---|---|
| *Total Sample* | 993 | 100.0 |
| Person without disability | 765 | 77.0 |
| Person with disability | 228 | 23.0 |
|  |  |  |
| *Types of Disabilities (Not Mutually Exclusive)* | 228 | 100.0 |
| Deaf/difficulty hearing | 33 | 14.5 |
| Blind/difficulty seeing | 10 | 4.4 |
| Difficulty concentrating/making decisions | 113 | 49.6 |
| Difficulty walking/climbing stairs | 57 | 25.0 |
| Difficulty dressing/bathing | 7 | 3.1 |
| Difficulty doing errands alone | 39 | 17.2 |
| Difficulty interacting with others | 47 | 20.6 |
| Long-term health impairment | 106 | 46.5 |
|  |  |  |
| *Difficulties Among People with Disabilities* | 228 | 100.0 |
| Health condition has affected my ability to complete work duties with moderate or severe difficulty | 32 | 14.0 |
| Have you disclosed your health condition, impairment, or disability to your employer? |  |  |
|   Yes | 117 | 51.8 |
|   No | 57 | 25.2 |
|   It's complicated | 52 | 23.0 |
|   Did not respond | 2 | 0.9 |
| I have sometimes been unfairly treated because of my health condition, impairment, or disability. | 34 | 14.9 |
| At work I feel socially isolated because of my health condition, impairment, or disability | 35 | 15.4 |
| I have not disclosed my health condition, impairment, or disability at work because I am afraid of being stigmatized | 38 | 16.7 |

Source: Authors' computations based on original survey.

Table 2. Descriptive Statistics for Job Satisfaction and Perceptions of Work Experiences, by Disability Status

|  | Disability Mean | No Disability Mean | Disability Gap (Difference in Means) | Disability Gap (Regression Adjusted) |
|---|---|---|---|---|
| Somewhat or very satisfied in job | 0.592 | 0.648 | -0.056 | -0.047 |
| Index of agreement on job autonomy | 0.575 | 0.599 | -0.023 | 0.015 |
| Index of turnover intentions | 0.420 | 0.323 | 0.096*** | 0.095*** |
| Index of employee organizational commitment | 0.459 | 0.530 | -0.071** | -0.056* |
| Index of employee organizational citizenship behaviors | 0.550 | 0.557 | -0.007 | 0.015 |
| Index of perceived organizational support | 0.309 | 0.403 | -0.094*** | -0.085** |
| Index of employer openness to differences | 0.453 | 0.561 | -0.107*** | -0.098*** |
| Index of climate for inclusion | 0.348 | 0.416 | -0.069** | -0.055* |
| Index of treatment of people with disabilities | 0.415 | 0.438 | -0.023 | -0.013 |
| Index of relationship with manager (leader-member exchange) | 0.657 | 0.761 | -0.105*** | -0.095*** |
| Index of relationships with coworkers (coworker exchange) | 0.706 | 0.801 | -0.095*** | -0.067*** |

Note: Sample size 993. *** statistically significant at 1%, ** at 5%, and * at 10% in 2-tail t tests. Figures denote the proportion of the sample who agree with the statements. See Appendix Table 3 for sample means using Disclosed Disability status, and for the detailed indicators included in the indices. The regression-adjusted disability gaps control for age, gender, race/ethnicity, marital status, education, income above $75,000, number of children at home, managerial role, full-time worker, and tenure at the employer.

Table 3. Descriptive Statistics for Work From Home Characteristics, by Disability Status

| Work From Home (WFH) Characteristics | Disability Mean | No Disability Mean | Difference |
|---|---|---|---|
| WFH before pandemic | 0.189 | 0.160 | 0.029 |
|    If no: did not WFH b/c not permitted | 0.454 | 0.443 | 0.010 |
|    If yes: WFH to coordinate w/ family schedule | 0.463 | 0.500 | -0.037 |
|    If yes: WFH 3 or more days/week | 0.439 | 0.333 | 0.106 |
| WFH currently | 0.529 | 0.529 | -0.001 |
|    If no: do not WFH b/c not permitted | 0.292 | 0.288 | 0.005 |
|    If yes: WFH to coordinate w/ family schedule | 0.331 | 0.371 | -0.040 |
|    If yes: WFH 3 or more days/week | 0.825 | 0.716 | 0.109** |
| I want to WFH more than I do now | 0.462 | 0.333 | 0.129** |
| Index of agreement that WFH had positive effects before pandemic | 0.544 | 0.506 | 0.039 |
| Index of agreement that WFH has positive effects currently | 0.651 | 0.628 | 0.023 |

Note: Sample size 993. *** statistically significant at 1%, ** at 5%, and * at 10% in 2-tail t tests. See Appendix Table 4 for sample means using Disclosed Disability status, and Appendix Table 5 for indicators included in the indices of agreement on WFH positive effects.

Table 4. Regression Results for Association Between Work Experiences, Disability, and Work from Home Frequency

| Variable | Job satisfaction | Job autonomy | Turnover intentions | Organizational commitment | Organizational citizenship behaviors | Perceived organizational support |
|---|---|---|---|---|---|---|
| Work from home (exclude: no WFH) | | | | | | |
| WFH < 3 days/week | 0.124*** | 0.226*** | -0.063 | 0.127*** | 0.080** | 0.188*** |
| | (0.047) | (0.034) | (0.041) | (0.042) | (0.038) | (0.041) |
| WFH 3+ days/week | 0.195*** | 0.294*** | -0.162*** | 0.176*** | 0.088*** | 0.253*** |
| | (0.041) | (0.029) | (0.036) | (0.036) | (0.033) | (0.035) |
| Disability interactions with: | | | | | | |
| No WFH | -0.080 | 0.040 | 0.159*** | -0.046 | 0.020 | -0.035 |
| | (0.054) | (0.039) | (0.047) | (0.048) | (0.043) | (0.047) |
| WFH < 3 days/week | -0.013 | -0.131** | 0.016 | -0.016 | -0.010 | -0.135* |
| | (0.087) | (0.062) | (0.076) | (0.077) | (0.069) | (0.076) |
| WFH 3+ days/week | -0.058 | -0.005 | 0.084 | -0.117** | -0.002 | -0.164*** |
| | (0.061) | (0.044) | (0.053) | (0.054) | (0.048) | (0.053) |

| Variable | Employer openness to difference | Climate for inclusion | Treatment of people with disabilities | Manager relations | Coworker relations |
|---|---|---|---|---|---|
| Work from home (exclude: no WFH) | | | | | |
| WFH < 3 days/week | 0.108*** | 0.122*** | 0.033 | 0.110*** | 0.010 |
| | (0.042) | (0.042) | (0.030) | (0.035) | (0.032) |
| WFH 3+ days/week | 0.154*** | 0.190*** | 0.042* | 0.124*** | 0.035 |
| | (0.036) | (0.036) | (0.025) | (0.030) | (0.027) |
| Disability interactions with: | | | | | |
| No WFH | -0.091* | -0.026 | -0.011 | -0.035 | -0.057 |
| | (0.048) | (0.048) | (0.034) | (0.040) | (0.036) |
| WFH < 3 days/week | -0.094 | 0.005 | -0.016 | -0.135** | -0.072 |
| | (0.077) | (0.077) | (0.054) | (0.064) | (0.059) |
| WFH 3+ days/week | -0.134** | -0.148*** | -0.019 | -0.171*** | -0.078* |
| | (0.054) | (0.054) | (0.038) | (0.044) | (0.041) |

Note: Sample size 993. *** statistically significant at 1%, ** at 5%, and * at 10% in 2-tail t tests. All regressions include controls for age, gender, race/ethnicity, marital status, education, income above $75,000, number of children at home, managerial role, full-time worker, and tenure at the employer.

Appendix Table 1. Sample Means for Demographic Characteristics, by Disability Status

| Demographic Characteristic | Disability Mean | No Disability Mean | Difference |
|---|---|---|---|
| Age | 43.638 | 44.935 | -1.297 |
| # Children at home | 1.772 | 2.050 | -0.278*** |
| Gender | | | |
|   Man | 0.085 | 0.108 | -0.023 |
|   Woman | 0.848 | 0.868 | -0.020 |
|   Nonbinary | 0.067 | 0.024 | 0.043*** |
| Race/Ethnicity | | | |
|   Black | 0.124 | 0.118 | 0.007 |
|   White | 0.836 | 0.837 | -0.002 |
|   Hispanic | 0.040 | 0.026 | 0.014 |
|   American Indian/Alaska Native | 0.058 | 0.034 | 0.023 |
|   Asian/Pacific Islander | 0.013 | 0.017 | -0.004 |
|   Multiracial/other | 0.089 | 0.052 | 0.037** |
| Married | 0.482 | 0.666 | -0.184*** |
| Education | | | |
|   <9th grade | 0.004 | 0.000 | 0.004* |
|   High school graduate | 0.098 | 0.069 | 0.029 |
|   Some college, no degree | 0.209 | 0.147 | 0.062** |
|   Associate degree | 0.204 | 0.167 | 0.038 |
|   Bachelor's degree | 0.311 | 0.371 | -0.060 |
|   Master's degree | 0.151 | 0.181 | -0.030 |
|   Professional degree/PhD | 0.022 | 0.065 | -0.043** |
| Income >=$75,000 | 0.223 | 0.364 | -0.141*** |
| Works full-time | 0.886 | 0.882 | 0.004 |
| Worked at employer before pandemic | 0.741 | 0.800 | -0.059* |
| Worked at employer <= 5 years | 0.513 | 0.400 | 0.113*** |
| Has management role | 0.088 | 0.125 | -0.037 |

Note: Sample size 993. *** statistically significant at 1%, ** at 5%, and * at 10% in 2-tail t tests. See Appendix Table 2 for sample means using Disclosed Disability status.

Appendix Table 2. Sample Means for Demographic Characteristics, by Disclosed Disability Status

| Demographic Characteristic | Disclosed Disability Mean | Undisclosed/ No Disability Mean | Difference |
|---|---|---|---|
| Age | 45.468 | 44.530 | 0.938 |
| # Children at home | 1.809 | 2.010 | -0.202* |
| Gender | | | |
|   Man | 0.079 | 0.106 | -0.027 |
|   Woman | 0.877 | 0.861 | 0.016 |
|   Nonbinary | 0.044 | 0.032 | 0.011 |
| Race/Ethnicity | | | |
|   Black | 0.112 | 0.120 | -0.008 |
|   White | 0.871 | 0.832 | 0.038 |
|   Hispanic | 0.052 | 0.027 | 0.025 |
|   American Indian/Alaska Native | 0.026 | 0.042 | -0.016 |
|   Asian/Pacific Islander | 0.009 | 0.017 | -0.009 |
|   Multiracial/other | 0.060 | 0.060 | 0.000 |
| Married | 0.482 | 0.643 | -0.161*** |
| Education | | | |
|   <9th grade | 0.000 | 0.001 | -0.001 |
|   High school graduate | 0.121 | 0.069 | 0.051** |
|   Some college, no degree | 0.207 | 0.155 | 0.052 |
|   Associate degree | 0.155 | 0.178 | -0.023 |
|   Bachelor's degree | 0.336 | 0.360 | -0.024 |
|   Master's degree | 0.147 | 0.178 | -0.032 |
|   Professional degree/PhD | 0.034 | 0.058 | -0.023 |
| Income >=$75,000 | 0.216 | 0.348 | -0.132*** |
| Works full-time | 0.872 | 0.885 | -0.013 |
| Worked at employer before pandemic | 0.726 | 0.794 | -0.068* |
| Worked at employer <= 5 years | 0.547 | 0.410 | 0.137*** |
| Has management role | 0.086 | 0.121 | -0.035 |

Note: Sample size 993. *** statistically significant at 1%, ** at 5%, and * at 10% in 2-tail t tests.

Appendix Table 3. Detailed Indicators and Sample Means for Job Satisfaction and Perceptions of Job Experiences, by Disability Status

| % of People Who Agree with These Statements: | Self-Reported Disb. Mean | No Disb. Mean | Disb. Diff. | Disclosed Disb. Mean | NoDisb/Undisc Mean | Disb. Diff. |
|---|---|---|---|---|---|---|
| **Somewhat or very satisfied in job** | **0.592** | **0.648** | **-0.056** | **0.590** | **0.641** | **-0.051** |
| **Index of agreement on job autonomy** | **0.575** | **0.599** | **-0.023** | **0.593** | **0.593** | **0.001** |
| Currently pandemic job allows me to decide when to begin and end work each day | 0.356 | 0.391 | -0.035 | 0.365 | 0.385 | -0.020 |
| Currently I have some control over the sequencing of my work activities | 0.704 | 0.744 | -0.040 | 0.707 | 0.738 | -0.031 |
| Currently I can decide when to do particular work activities | 0.665 | 0.665 | 0.000 | 0.707 | 0.659 | 0.048 |
| **Index of turnover intentions** | **0.420** | **0.323** | **0.096***** | **0.407** | **0.337** | **0.070*** |
| Currently I plan to look for job outside this company during the next year | 0.363 | 0.283 | 0.080** | 0.359 | 0.294 | 0.065 |
| Currently I often think about quitting my job at this company | 0.439 | 0.334 | 0.104*** | 0.419 | 0.350 | 0.069 |
| Currently I want to get a new job | 0.461 | 0.353 | 0.108*** | 0.444 | 0.368 | 0.076 |
| **Index of organizational commitment** | **0.459** | **0.530** | **-0.071**** | **0.481** | **0.518** | **-0.037** |
| I feel a strong sense of 'belonging' to the employer | 0.469 | 0.577 | -0.108*** | 0.470 | 0.563 | -0.093* |
| I feel like 'part of the family' at the employer | 0.427 | 0.526 | -0.099*** | 0.436 | 0.513 | -0.077 |
| The employer has a great deal of personal meaning for me | 0.476 | 0.487 | -0.011 | 0.538 | 0.477 | 0.061 |
| **Index of organizational citizenship behaviors** | **0.550** | **0.557** | **-0.007** | **0.584** | **0.551** | **0.033** |
| I keep up with developments at the employer | 0.610 | 0.634 | -0.024 | 0.615 | 0.630 | -0.014 |
| I offer ideas to improve the functioning of the employer | 0.430 | 0.450 | -0.020 | 0.453 | 0.444 | 0.009 |
| I take action to protect the employer from potential problems | 0.610 | 0.588 | 0.022 | 0.684 | 0.581 | 0.103** |
| **Index of perceived organizational support** | **0.309** | **0.403** | **-0.094**** | **0.376** | **0.383** | **-0.006** |
| The employer really cares about my well-being | 0.339 | 0.433 | -0.093** | 0.371 | 0.417 | -0.046 |
| The employer takes pride in my accomplishments at work | 0.313 | 0.412 | -0.100*** | 0.405 | 0.387 | 0.018 |
| The employer cares about my opinions | 0.276 | 0.366 | -0.090** | 0.353 | 0.344 | 0.009 |

Continued on next page

Appendix Table 3 Continued. Detailed Indicators and Sample Means for Job Satisfaction and Perceptions of Job Experiences, by Disability Status

| % of People Who Agree with These Statements: | Self-Reported Disb. Mean | No Disb. Mean | Disb. Diff. | Disclosed Disb. Mean | NoDisb/Undisc Mean | Disb. Diff. |
|---|---|---|---|---|---|---|
| **Index of employer openness to differences** | **0.453** | **0.561** | **-0.107***** | **0.490** | **0.542** | **-0.052** |
| The employer has a non-threatening environment in which people can reveal their 'true' selves | 0.482 | 0.616 | -0.133*** | 0.521 | 0.594 | -0.072 |
| Employees are valued for who they are as people, not just for the jobs that they fill | 0.355 | 0.450 | -0.094** | 0.402 | 0.432 | -0.030 |
| We have a culture in which employees appreciate the differences that people bring to the workplace | 0.522 | 0.616 | -0.095** | 0.547 | 0.601 | -0.054 |
| **Index of climate for inclusion** | **0.348** | **0.416** | **-0.069**** | **0.385** | **0.403** | **-0.018** |
| Employee input is actively sought | 0.390 | 0.463 | -0.072* | 0.402 | 0.452 | -0.050 |
| Everyone's opinions for how to do things better are given serious consideration | 0.298 | 0.343 | -0.045 | 0.333 | 0.333 | 0.001 |
| Employees' insights are used to rethink or redefine work practices | 0.325 | 0.399 | -0.075** | 0.359 | 0.385 | -0.026 |
| Management exercises the belief that problem-solving is improved when input from different roles, ranks, and functions is considered | 0.377 | 0.463 | -0.086** | 0.444 | 0.443 | 0.001 |
| **Index of treatment of people with disabilities** | **0.415** | **0.438** | **-0.023** | **0.459** | **0.429** | **0.030** |
| Employees with disabilities have the same opportunities as people without disabilities | 0.469 | 0.496 | -0.027 | 0.521 | 0.486 | 0.036 |
| The employer is making strong efforts to improve conditions and opportunities for people with disabilities | 0.382 | 0.439 | -0.058 | 0.419 | 0.427 | -0.008 |
| The environment is such that, when accommodations are made, people with disabilities can be just as productive as people without disabilities | 0.447 | 0.489 | -0.041 | 0.513 | 0.475 | 0.038 |
| Bias against people with disabilities exists where I work | 0.259 | 0.122 | 0.136*** | 0.291 | 0.135 | 0.155*** |
| Where I work, employees without disabilities are treated better than employees with disabilities | 0.137 | 0.067 | 0.070*** | 0.162 | 0.072 | 0.090*** |
| Top management commits to hire people with disabilities | 0.138 | 0.174 | -0.036 | 0.148 | 0.168 | -0.021 |
| Employees treat people with disabilities with respect | 0.586 | 0.679 | -0.093*** | 0.624 | 0.662 | -0.039 |
| My manager treats people with disabilities with respect | 0.659 | 0.702 | -0.042 | 0.707 | 0.690 | 0.017 |
| The employer is responsive to the needs of people with disabilities | 0.471 | 0.532 | -0.061 | 0.513 | 0.519 | -0.006 |
| My manager is responsive to the needs of people with disabilities | 0.608 | 0.672 | -0.064* | 0.692 | 0.652 | 0.040 |

Appendix Table 3 Continued. Detailed Indicators and Sample Means for Job Satisfaction and Perceptions of Job Experiences, by Disability Status

| % of People Who Agree with These Statements: | Self-Reported Disb. Mean | No Disb. Mean | Disb. Diff. | Disclosed Disb. Mean | NoDisb/Undisc Mean | Disb. Diff. |
|---|---|---|---|---|---|---|
| **Index of manager relations (leader-member exchange)** | **0.657** | **0.761** | **-0.105***** | **0.706** | **0.742** | **-0.036** |
| I usually know how satisfied my manager is with what I do | 0.699 | 0.801 | -0.101*** | 0.726 | 0.784 | -0.058 |
| I feel that my manager understands my problems and needs | 0.562 | 0.712 | -0.150*** | 0.615 | 0.686 | -0.070 |
| I feel that my manager recognizes my potential | 0.611 | 0.750 | -0.140*** | 0.650 | 0.728 | -0.078* |
| I can count on my manager to support me even when I'm in a tough situation at work | 0.668 | 0.755 | -0.086*** | 0.718 | 0.737 | -0.019 |
| I would support my manager's decision even if he or she was not present | 0.752 | 0.801 | -0.049 | 0.821 | 0.786 | 0.035 |
| I have an effective working relationship with my manager | 0.730 | 0.821 | -0.091*** | 0.769 | 0.804 | -0.035 |
| If necessary, my manager would use his or her power and influence to help me | 0.575 | 0.692 | -0.117*** | 0.641 | 0.669 | -0.028 |
| **Index of coworker relations (coworker exchange)** | **0.706** | **0.801** | **-0.095**** | **0.698** | **0.790** | **-0.092**** |
| I usually know how satisfied my coworkers are with what I do | 0.681 | 0.804 | -0.123*** | 0.672 | 0.790 | -0.117*** |
| I feel that my coworkers understand my problems and needs | 0.590 | 0.711 | -0.121*** | 0.595 | 0.695 | -0.100** |
| I can count on my coworkers to support me even when I'm in a tough situation at work | 0.740 | 0.818 | -0.078*** | 0.724 | 0.811 | -0.086** |
| I would support my coworkers' decisions even if they were not present | 0.811 | 0.837 | -0.027 | 0.810 | 0.834 | -0.024 |
| I have an effective working relationship with my coworkers | 0.828 | 0.893 | -0.064*** | 0.802 | 0.888 | -0.086*** |
| If necessary, my coworkers would use their power and influence to help me | 0.586 | 0.744 | -0.158*** | 0.586 | 0.724 | -0.138*** |

Note: Sample size 993. *** statistically significant at 1%, ** at 5%, and * at 10% in 2-tail t tests. Figures denote % who agree with the statements included in each index. Indicators in each index are weighted equally.

Appendix Table 4. Detailed Indicators and Sample Means by Disability Status for Work From Home, Before Pandemic and Currently

| Work From Home (WFH) Characteristics | Self-Reported Disb. Mean | No Disb. Mean | Disb. Diff. | Disclosed Disb. Mean | NoDisb/Undisc Mean | Disb. Diff. |
|---|---|---|---|---|---|---|
| **Index of agreement that WFH had positive effects before pandemic** | **0.544** | **0.506** | **0.039** | **0.534** | **0.513** | **0.022** |
| WFH before pandemic had pos effect on morale | 0.825 | 0.805 | 0.020 | 0.810 | 0.810 | -0.001 |
| WFH before pandemic had pos effect on staying at employer | 0.732 | 0.658 | 0.073 | 0.714 | 0.671 | 0.043 |
| WFH before pandemic had pos effect on relations with coworkers | 0.293 | 0.258 | 0.034 | 0.286 | 0.264 | 0.021 |
| WFH before pandemic had pos effect on acquiring new skills | 0.195 | 0.183 | 0.012 | 0.143 | 0.193 | -0.050 |
| WFH before pandemic had pos effect on stress at work | 0.659 | 0.675 | -0.016 | 0.667 | 0.671 | -0.005 |
| WFH before pandemic had pos effect on productivity | 0.707 | 0.625 | 0.082 | 0.714 | 0.636 | 0.079 |
| WFH before pandemic had pos effect on ability to be promoted | 0.098 | 0.042 | 0.056 | 0.143 | 0.043 | 0.100* |
| WFH before pandemic had pos effect on ability to care for family | 0.805 | 0.689 | 0.116 | 0.762 | 0.712 | 0.050 |
| WFH before pandemic had pos effect on ability to manage health | 0.585 | 0.633 | -0.048 | 0.571 | 0.629 | -0.057 |
| **Index of agreement that WFH has positive effects currently** | **0.651** | **0.628** | **0.023** | **0.720** | **0.623** | **0.097**** |
| WFH currently has pos effect on morale | 0.840 | 0.822 | 0.018 | 0.912 | 0.816 | 0.096* |
| WFH currently has pos effect on staying at employer | 0.767 | 0.741 | 0.026 | 0.842 | 0.735 | 0.107* |
| WFH currently has pos effect on relations with coworkers | 0.466 | 0.378 | 0.088* | 0.564 | 0.378 | 0.185*** |
| WFH currently has pos effect on acquiring new skills | 0.445 | 0.423 | 0.022 | 0.446 | 0.426 | 0.021 |
| WFH currently has pos effect on stress at work | 0.750 | 0.777 | -0.027 | 0.807 | 0.766 | 0.041 |
| WFH currently has pos effect on productivity | 0.758 | 0.751 | 0.008 | 0.860 | 0.739 | 0.120** |
| WFH currently has pos effect on ability to be promoted | 0.160 | 0.151 | 0.009 | 0.214 | 0.146 | 0.069 |
| WFH currently has pos effect on ability to care for family | 0.824 | 0.814 | 0.009 | 0.860 | 0.811 | 0.048 |
| WFH currently has pos effect on ability to manage health | 0.783 | 0.767 | 0.016 | 0.807 | 0.767 | 0.040 |

Note: Sample size 993. *** statistically significant at 1%, ** at 5%, and * at 10% in 2-tail t tests. All figures denote % who agree with the statement. Indicators in each index are weighted equally.

Appendix Table 5. Sample Means for Work From Home Characteristics, by Disclosed Disability Status

| Work From Home (WFH) Characteristics | Disclosed Disability Mean | Undisclosed/ No Disability Mean | Difference |
|---|---|---|---|
| WFH before pandemic | 0.188 | 0.163 | 0.025 |
|    If no: did not WFH b/c not permitted | 0.400 | 0.451 | -0.051 |
|    If yes: WFH to coordinate w/ family schedule | 0.381 | 0.507 | -0.126 |
|    If yes: WFH 2 or more days/week | 0.476 | 0.343 | 0.133 |
| WFH currently | 0.487 | 0.535 | -0.048 |
|    If no: do not WFH b/c not permitted | 0.300 | 0.287 | 0.013 |
|    If yes: WFH to coordinate w/ family schedule | 0.386 | 0.359 | 0.027 |
|    If yes: WFH 2 or more days/week | 0.842 | 0.729 | 0.113* |
| I want to WFH more than I do now | 0.456 | 0.351 | 0.105 |
| Index of agreement on WFH positive effects before pandemic | 0.534 | 0.513 | 0.022 |
| Index of agreement on WFH positive effects currently | 0.720 | 0.623 | 0.097** |

Note: Sample size 993. *** statistically significant at 1%, ** at 5%, and * at 10% in 2-tail t tests. See Appendix Table 4 for indicators included in the indices of agreement on WFH positive effects.

Appendix Table 6. Regression Results for Association Between Work Experiences, Disability, and Reason to Work from Home

| Variable | Job satisfaction | Job autonomy | Turnover intentions | Organizational commitment | Organizational citizenship behaviors | Perceived organizational support |
|---|---|---|---|---|---|---|
| Work from home (exclude: no WFH) | | | | | | |
| WFH just for pandemic or employer | 0.136*** | 0.258*** | -0.091** | 0.135*** | 0.074** | 0.211*** |
| | (0.046) | (0.033) | (0.040) | (0.041) | (0.037) | (0.040) |
| WFH for benefit of employee | 0.191*** | 0.277*** | -0.147*** | 0.173*** | 0.092*** | 0.240*** |
| | (0.041) | (0.030) | (0.036) | (0.037) | (0.033) | (0.036) |
| Disability interactions with: | | | | | | |
| No WFH | -0.080 | 0.040 | 0.159*** | -0.046 | 0.020 | -0.034 |
| | (0.054) | (0.039) | (0.048) | (0.048) | (0.043) | (0.047) |
| WFH just for pandemic or employer | -0.044 | -0.028 | 0.026 | -0.004 | 0.046 | -0.100 |
| | (0.083) | (0.060) | (0.073) | (0.073) | (0.066) | (0.072) |
| WFH for benefit of employee | -0.043 | -0.046 | 0.077 | -0.128** | -0.033 | -0.181*** |
| | (0.063) | (0.045) | (0.055) | (0.056) | (0.050) | (0.055) |

Appendix Table 6 Continued.  Regression Results for Association Between Work Experiences, Disability, and Reason to Work From Home

| Variable | Employer openness to difference | Climate for inclusion | Treatment of people with disabilities | Manager relations | Coworker relations |
|---|---|---|---|---|---|
| Work from home (excl.: no WFH) | | | | | |
|   WFH just for pandemic or employer | 0.102** | 0.155*** | 0.016 | 0.117*** | 0.023 |
| | (0.040) | (0.041) | (0.029) | (0.034) | (0.031) |
|   WFH for benefit of employee | 0.161*** | 0.170*** | 0.054** | 0.120*** | 0.027 |
| | (0.036) | (0.036) | (0.026) | (0.030) | (0.028) |
| Disability interactions with: | | | | | |
|   No WFH | -0.091* | -0.026 | -0.011 | -0.035 | -0.057 |
| | (0.048) | (0.048) | (0.034) | (0.040) | (0.036) |
|   WFH just for pandemic or employer | 0.005 | -0.061 | 0.026 | -0.207*** | -0.015 |
| | (0.073) | (0.073) | (0.052) | (0.061) | (0.055) |
|   WFH for benefit of employee | -0.190*** | -0.116** | -0.045 | -0.134*** | -0.106** |
| | (0.055) | (0.055) | (0.039) | (0.046) | (0.042) |

Note: Sample size 993. *** statistically significant at 1%, ** at 5%, and * at 10% in 2-tail t tests. All regressions include control variables for age, gender, race/ethnicity, married, education, income above $75,000, number of children at home, managerial role, full-time worker, and tenure at the employer.

# Appendix Notes:
## Sample, Disclosed Disabilities, and Perceptions of Work Experience Scales

**Sample:**

Our survey instrument included questions on employees' awareness and perceptions of employer policies that address the physical and mental health needs of workers. It focused on employer practices around work from home, and it also asked about work experiences before the pandemic. Survey questions used the wording "Before March 2020" to denote the period before the pandemic started, and phrases such as "currently" or "today" to denote the current period at the time of the survey. We used the data to calculate simple summary statistics on the prevalence of work from home, disability disclosure, perceptions of workplace inclusiveness, treatment of people with disabilities, and various measures of job satisfaction.

In collaboration with partners from the healthcare system, our research team distributed the survey link via Qualtrics to employees. The distribution included a cover letter providing details about the study and outlining the informed consent process.

Qualtrics reported 1,405 respondents. We dropped 135 of those respondents because they clicked on the survey link but did not answer any questions. An additional 277 respondents did not respond to questions about their disability status, so we also dropped these individuals, leaving a sample of 993. Robustness checks in which these 277 respondents were kept in the sample and assumed to have no disability yielded substantively similar results to those reported in the paper.

The eight questions used to assess disability are:
- "Are you deaf or do you have serious difficulty hearing?"
- "Are you blind or do you have serious difficulty seeing even when wearing glasses?"
- "Do you have serious difficulty concentrating, remembering, or making decisions?"
- "Do you have serious difficulty walking or climbing stairs?"
- "Do you have difficulty dressing or bathing?"
- "Do you have difficulty doing errands alone such as visiting a doctor's office or shopping?"
- "Do you have difficulty interacting and/or communicating with others??"
- "Do you have a long-term health problem or impairment that limits the kind or amount of work, housework, school, parenting, recreation, or other activities you can do?"

**Discussion of Sample Means:**

Sample means for people with and without disabilities show meaningful differences in the number of children, education, marital status, and income (Appendix Table 1). In particular, individuals with disabilities are less likely to have a college degree or be married compared to people without disabilities. People with disabilities also have fewer children and lower income on average compared to people without disabilities. These findings on key demographic indicators among people with and without disabilities are consistent with previous studies. Sample means further show that workers with disabilities were less likely to have worked at the employer before the pandemic and were more likely to have a relatively short tenure (five years or less). However, there are no statistically significant differences between people with and without disabilities in the likelihood of working full-time or having a management role. Overall, the results for people with disclosed disabilities are comparable (Appendix Table 2), except that the differences by disability status in being multiracial/other race and having a professional degree or PhD are no longer statistically significant.

**Disclosed Disabilities:**

The difference between people with and without disclosed disabilities is particularly large for the perceived positive effect on coworker relations. It may seem counter-intuitive that working from home is good for coworker relations, but this result could be driven by increased control over the work environment and the removal of barriers related to physical mobility. Employees may be better able to control volume and implement technologies like on-screen closed captions in meetings and online scheduling in ways that allow them to have more mental and emotional preparation and/or resources available for engaging with colleagues. Those factors could enhance the job satisfaction that comes with work from home for many, albeit not all, employees. Hence working from home may allow some people with disabilities to tailor their working interactions to meet their needs, especially if they have disclosed their disability to their employer.

Overall, these results indicate that the index of agreement on the benefits of remote work is the highest for people who have disclosed their disability to the employer (0.720), and this index is substantially higher than it is for people without disclosed disabilities (0.623) and even for all people with disabilities regardless of disclosure (0.651). Also of note, all of the detailed indicators used to construct the indices of agreement have a higher score for respondents' current views compared to their views of work from home before the pandemic. This set of results may reflect remote work as an *unintentional accommodation* during the pandemic for people with unknown, undisclosed, or invisible disabilities (Cohen and Rodgers 2024). In particular, when they allowed remote work during the pandemic, employers may not even have known that they were providing accommodations for employees with a disability. This possibility is supported by evidence indicating that before the pandemic, accommodation requests from workers with mental health impairments were less likely to be seen as reasonable and granted by employers compared to requests from people with physical disabilities.

Results for organizational commitment are muted for people who have disclosed their disabilities to their employer. In most cases, the disability gap in these job experience indicators is smaller in magnitude, and often no longer statistically significant, when we focus on people with disclosed disabilities. On the one hand, one might imagine that those who disclose their disabilities automatically have more equitable opportunities through changes in employer expectations about their work performance. On the other hand, more equitable opportunities are not an automatic byproduct of disclosure; rather, they are a product of management choosing to grant accommodation requests or not. It is also possible that those who chose to disclose had better relations with managers before disclosure.

In many cases the disability differences in perceptions of how people with disabilities are treated are lessened when we consider people with disclosed disabilities. There are only two instances in which the gap is even larger and highly statistically significant: people with disclosed disabilities are even more likely to state that their workplace has a bias against people with disabilities and that employees without disabilities are treated better than employees with disabilities. We are thus particularly interested in the extent to which working from home can help to mitigate these negative perceptions of workplace bias against people with disabilities.

When it comes to relationships with coworkers and managers, once we restrict the disability sample to individuals with disclosed disabilities, we see that the experiences of people who have disclosed disabilities are closer to those without disabilities in relationships with managers but are consistently negative with respect to coworkers.

**Perceptions of Work Experience Scales:**

Perceived organizational support:

a. The organization really cares about my well-being.
   b. The organization takes pride in my accomplishments at work.
   c. The organization cares about my opinions.

Cites:
3 items from Wayne et al. based on longer scale from Eisenberger et al.
Wayne, S., Shore, L., & Liden, R. (1997). Perceived organizational support and leader-member exchange: A social exchange perspective. *Academy of management Journal, 40,* 82-111.
Eisenberger, R., Huntington, R., Hutchison, S., & Sowa, D. (1986). Perceived organizational support. *Journal of Applied Psychology, 71,* 500-507.

Organizational commitment:

   This is known more specifically as "affective organizational commitment"
   a. I feel a strong sense of "belonging" to my organization.
   b. I feel like "part of the family" at my organization.
   c. My organization has a great deal of personal meaning for me.

Cite:
Drawn from Meyer, J.P, Allen, M.J., & Smith, C.A. (1993). Commitment to organizations and occupations: Extension and test of a three-component conceptualization. *Journal of Applied Psychology, 78,* 538-551.

Organizational citizenship behaviors:

   How often do you engage in these behaviors?
   a. Keep up with developments in the organization.
   b. Offer ideas to improve the functioning of the organization.
   c. Take action to protect the organization from potential problems.

Cite:
Lee, K. and Allen, N.J. (2002), "Organizational citizenship behavior and workplace deviance: the role of affect and cognitions", Journal of Applied Psychology, Vol. 87 No. 1, pp. 131-142, doi: 10.1037// 0021-9010.87.1.131.

Leader-member exchange:

   Please indicate the extent to which you agree with these statements about the relationship between you and your supervisor/manager? *(Please circle ONE answer for each item)*

   I usually know how satisfied my manager is with what I do.
   I feel that my manager understands my problems and needs
   I feel that my manager recognizes my potential.
   I can count on my manager to support me even when I'm in a tough situation at work.
   I would support my manager's decisions even if he or she was not present.
   I have an effective working relationship with my manager.
   If necessary, my manager would use his or her power and influence to help me.

Adapted from:
Graen, G. B., & Uhl-Bien, M. (1995). Relationship-based approach to leadership: Development of leader-member exchange (LMX) theory of leadership over 25 years: Applying a multi-level multi-domain perspective. *The leadership quarterly*, *6*(2), 219-247.

Coworker exchange:

Please indicate the extent to which you agree with these statements about the relationship between you and your coworkers (Please circle ONE answer for each item) ※

E34. I usually know how satisfied my coworkers are with what I do.
E35. I feel that my coworkers understand my problems and needs.
E36. I can count on my coworkers to support me even when I'm in a tough situation at work.
E37. I would support my coworkers' decisions even if they were not present.
E38. I have an effective working relationship with my coworkers.
E39. If necessary, my coworkers would use their power and influence to help me.

Cite:
Sherony, K. M., & Green, S. G. (2002). Coworker exchange: relationships between coworkers, leader-member exchange, and work attitudes. *Journal of applied psychology*, *87*(3), 542.

Climate for inclusion:

This is a subscale representing the dimension of climate for inclusion in decision-making

How would you rate the inclusiveness of your organization with regard to employees' ideas and experiences in general, particularly on each of the following? *Please CHOOSE ONE answer for each item.*

E13. In my organization, employee input is actively sought.
E14. In my organization, everyone's ideas for how to do things better are given serious consideration.
E15. In my organization, employees' insights are used to rethink or redefine work practices.
E16. Management exercises the belief that problem-solving is improved when input from different roles, ranks, and functions is considered.

Cite:
Nishii, L. H. (2013). The benefits of climate for inclusion for gender-diverse groups. *Academy of Management journal*, *56*(6), 1754-1774.

Turnover intentions:

    C9. I will look for a job outside this company during the next year.
    C10. I often think about quitting my job at this company.
    C11. I would like to get a new job.

Cite:
Konovsky, M. A., & Cropanzano, R. (1991). Perceived fairness of employee drug testing as a predictor of employee attitudes and job performance. *Journal of applied psychology*, *76*(5), 698.

Autonomy:

    C5. The job denies me any chance to use my personal initiative or judgment in carrying out the work.
    C6. The job gives me considerable opportunity for independence and freedom in how I do the work.
    C7. The job gives me considerable flexibility to work at my personal "peak" times (i.e., the times of day I feel most productive)
    C8. I have complete freedom to schedule my own work hours.

These four are all from Desroisers (2001), drawing from Hackman and Oldham (1978) and Breaugh (1985).

Cites:
Breaugh, J. A. (1985). The measurement of work autonomy. *Human relations*, *38*(6), 551-570.
Desrosiers, E. I. (2001). *Telework and work attitudes: The relationship between telecommuting and employee job satisfaction, organizational commitment, perceived organizational support, and perceived co-worker support*. Purdue University.
Hackman, J. R., & Oldham, G. R. (1975). Development of the job diagnostic survey. *Journal of Applied psychology*, *60*(2), 159.

The accommodations questions are all from:
Schur, L., Nishii, L., Adya, M., Kruse, D., Bruyère, S. M., & Blanck, P. (2014). Accommodating employees with and without disabilities. *Human Resource Management*, *53*(4), 593-621.